\documentclass[structabstract]{aa}
\usepackage{graphicx}
\usepackage{txfonts}
\usepackage{natbib}
\bibpunct{(}{)}{;}{a}{}{,}

\begin{document}

\title{Optimizing exoplanet transit searches around low-mass stars with
inclination constraints} 

\author{E.~Herrero\inst{\ref{inst1},\ref{inst2}}\and I.~Ribas\inst{\ref{inst1}} \and 
C.~Jordi\inst{\ref{inst2}} 
\and E.~F.~Guinan\inst{\ref{inst3}} \and S.~G.~Engle\inst{\ref{inst3}}
} 

\institute{Institut de Ci\`{e}ncies de l'Espai (CSIC-IEEC), Campus UAB, 
Facultat de Ci\`{e}ncies, Torre C5 parell, 2a pl, 08193 Bellaterra, 
Spain, \email{eherrero@ice.cat, iribas@ice.cat}\label{inst1}
\and
Dept. d'Astronomia i Meteorologia, Institut de Ci\`{e}ncies del Cosmos (ICC)
%\footnote{Associated with the Instituto de Ciencias del Espacio. Consejo Superior de Investigaciones Científicas.}
,Universitat de Barcelona (IEEC-UB), Mart\'{i} Franqu\`{e}s 1, E08028 Barcelona, Spain 
\email{carme.jordi@ub.edu}\label{inst2}
\and
Department of Astronomy \& Astrophysics, Villanova University, 800 Lancaster Avenue, Villanova, PA 19085, 
\email{edward.guinan@villanova.edu, sengle01@villanova.edu}\label{inst3}
}
\date{Received <date> /
Accepted <date>}

\abstract {} {We investigate a method to increase the efficiency of a targeted exoplanet
search with the transit technique by preselecting a subset of candidates from
large catalogs of stars. Assuming spin-orbit alignment, this can be
done by considering stars that have higher probability to be oriented nearly equator-on (inclination
close to $90^{\circ}$).} {We use activity-rotation velocity relations for
low-mass stars with a convective envelope to study the dependence of the 
position in the activity-$v\sin i$ diagram on the stellar axis inclination. We compose a
catalog of G-, K-, M-type main sequence simulated stars using
isochrones, an isotropic inclination distribution and empirical relations to
obtain their rotation periods and activity indexes. Then the activity - $v\sin
i$ diagram is filled and statistics are applied to trace the
areas containing the higher ratio of stars with inclinations above
$80^{\circ}$. A similar statistics is applied to stars from real catalogs with
$\log(R'_{HK})$ and $v\sin i$ data to find their probability of being equator-on.} {We present the method used to generate
the simulated star catalog and the subsequent statistics to find the highly
inclined stars from real catalogs using the activity-$v\sin i$ diagram. Several catalogs from the
literature are analysed and a subsample of stars with the highest probability
of being equator-on is presented.} { Assuming spin-orbit alignment, the efficiency of an exoplanet transit
search in the resulting subsample of probably highly inclined stars is estimated to be
two to three times higher than with a global search with no preselection.} 

\maketitle

\section{Introduction}

In the past years, the great development in the exoplanet detection techniques
and the new space missions and ground-based instrumentation is yielding a high
rate of new discoveries. Transiting planets represent a treasure in exoplanet 
research as they give us the possibility to determine both their size and
mass (the latter with a radial velocity follow-up), and also to study in detail
the properties of their atmospheres \citep{2010arXiv1012.2157L}. The stellar
light filtered through the planet's atmosphere during a transit allows to
obtain the transmission spectrum \citep{2002ApJ...568..377C, 2007Natur.448..169T}, whereas the
planet's dayside emission spectrum can be obtained during secondary eclipses,
yielding measurements of the composition and thermal structure of the 
planet's atmosphere among other properties 
\citep{2005ApJ...625L.135B,2006ApJ...650.1140B,2008Natur.456..767G}. Exoplanet
searches using the transit technique are nowadays providing a large number of
new findings. Since the first observation of a transit for the planet HD
209458 b \citep{2000ApJ...529L..45C,2000ApJ...529L..41H}, 133 transiting
exoplanets\footnote{http://exoplanet.eu/} have been detected and confirmed to date, 
more than 1200 candidates from the Kepler mission are awaiting confirmation
\citep{2011ApJ...728..117B,2011arXiv1102.0541B} and more discoveries are
expected from several ongoing surveys. Detecting an Earth-like planet
transiting a Sun-like bright star is one of the main objectives of exoplanet
research today.

Most exoplanet transit detection programs that are currently underway are
focused on large catalogs of stars with no pre-selection, basically
performing photometry of every possible target up to a certain limiting
magnitude. This necessarily makes such surveys quite inefficient, since
large amounts of data are processed for a relatively low transiting planet
yield. However, some stellar properties could be used to select
which stars may stand the best chances for finding transiting planets.  This 
can be especially important for some space missions, such as CoRot and Kepler (and 
future missions like PLATO and TESS), which must pre-select the targets and only 
downlink the data from the specific pixels containing the stars of interest. One
possible way  to perform a pre-selection is a metallicity-biased survey. In the case of solar-type stars it is
known that planet existence is strongly correlated with the presence of heavy
elements in the host star \citep[][e.g.,]{2005PThPS.158...24M}. Stellar
spectral type and age could also be considered in the case of searches of
terrestrial planets in the so called 'Habitable Zone -- HZ' 
\citep{2010AsBio..10..103K}, which is defined as the region around a star where one would expect
conditions for the existence of liquid water on a planet's surface. This
depends on the stellar luminosity, and so does the period of the planets
orbiting in the HZ. An example of a targeted survey with a pre-selection of
low-mass stars is the MEarth project \citep{2008AAS...212.4402C}. While the 
preselection increases the probability of detecting transits, limiting the 
target list to given stellar properties can result in the
introduction of selection effects on the properties of the exoplanets to be
discovered that need to be properly accounted for.

Considering a general targeted search of transiting planets, little effort has
been made in the past to put constraints to the input star catalogs with the
goal of increasing the transit detection rate. Relevant ideas were
presented by \cite{2010ApJ...712.1433B}: transit probabilities can be enhanced
if we are able to constrain the inclination of the stellar axes, and this can
significantly lower the number of targets to be observed in a transit
survey. A targeted transit search would imply to observe only a specific sample
of bright stars, which would be spread over the entire sky. This would require
several observatories at different latitudes and, in principle, a significant
amount of telescope time, as one star would be observed at a time. The
detection of transits or even the discovery of a habitable Earth-analog with
this approach is unrealistic.  However, if stellar inclination can be estimated
for a large sample of bright stars and expecting planets to orbit close to the
stellar equator plane, we could select and observe only those stars that have higher
probability to be equator-on, and therefore lower the number of targets and
increase the transit detection probability. In an isotropic distribution of the
stellar rotation axes, only about 17\% of the stars would have spin axis
inclinations above 80$^{\circ}$.

\cite{2010ApJ...712.1433B} calculate how constraining stellar
inclination affects the transit probabilities and the reduction of the
number of targets that need to be observed for a certain number of findings.
They also discuss some ways to measure stellar inclination, but this seems to
be the limiting point for the application of this approach. Perhaps, the
most plausible possibility is to obtain stellar inclinations from spectroscopic
$v\sin i$ measurements. For this, the true rotational period of the star could
be determined from photometric modulations caused by spots
\citep{2003csss...12..941M,2007ApJ...668L.163L} or modulation of the Ca II H
and K emission fluxes \citep{1984ApJ...279..763N}, and the stellar radius could
be obtained from stellar models. The major shortcoming of this method arises
from the uncertainties in all the ingredients: $v\sin i$, stellar rotational
period and stellar radius. Moreover, the behavior of the sine function itself
becomes a drawback, as it is weighed towards $\sin i=1$, thus changing very
slowly near $i=90^{\circ}$. All this makes necessary to know the observed
quantities to better than 1\% accuracy if one wants to select the stars with
$i>82^{\circ}$ \citep{1985PASP...97...57S}. With the currently available
techniques, it is not feasible to measure $\sin i$ to better than 10\%. This
approach would also be very time-intensive, requiring high resolution
spectroscopy and long time-series photometry. Therefore, it is unfeasible to
obtain a relatively large catalog of stellar inclination data.

Another approach is discussed in this paper, where we constrain the
inclination through the relation between the activity and the projected
rotational velocity of the star. Section 2 presents the principles of the
approach. In Sect. \ref{simul} we describe the simulation of large samples of
stars used to better understand this relation and to determine the probability
for an observed star to have its spin axis inclination above a specific angle.
This allows us to select the best candidates for a targeted exoplanet transit
search. Measurements of an activity indicator, such as $\log(R'_{HK})$, and of
$v\sin i$ are necessary to constrain stellar inclinations by this method.
A catalog selection is described in Sect. \ref{data} together with
the implementation of the selection method and the compilation of a subset of
stars that are expected to have inclinations close to 90$^{\circ}$. The
discussion included in Sect. \ref{discus} presents some comments on the
applicability of this preselection method and the complementarity of the
subsequent targeted transit search with the currently ongoing exoplanet
searching methods.

\section{Stellar inclination from activity and rotation}
\label{method}

A feasible approach to estimate stellar axis inclinations with the currently
available data is to exploit the existing relation between stellar activity and
rotation for main sequence late-type stars
\citep{1985AJ.....90.2103S,2003A&A...397..147P,2007AcA....57..149K}. In these
objects, the regime of differential rotation at the convective envelope plays a
key role at the generation of the magnetic fields through the dynamo effect.
These magnetic fields are essentially responsible for all the phenomena that
are globally known as stellar activity, and are also thought to be the main
rotation braking mechanism due to angular momentum loss through interaction
with the stellar wind. Essentially, the stellar mass (or spectral type), which
is related to the depth of the convective layer, and the rotational period
determine the amount of stellar activity \citep{1999A&A...344..911K}. This can
be measured through several indicators
\citep{1985AJ.....90.2103S,1998A&A...338..623M,1995A&A...294..515H}. Strong
evidence exists for an activity-rotation relation extending from solar-type
stars to less-massive dwarfs \citep{2003A&A...397..147P,2007AcA....57..149K}.

If an activity indicator like $\log(R'_{HK})$ can be compared to the projected
rotational velocity for a large sample of stars \citep{2011arXiv1103.0584J},
then those stars with $i\approx 90^{\rm o}$ ($\sin i\approx 1$) will have the
largest $v\sin i$ values for a certain activity, covering a specific area on
the activity-$v\sin i$ diagram for each spectral type. Studying the
relationship between chromospheric activity and projected rotational velocity through such
a diagram, statistics can be performed to find the area containing
preferentially stars with $i\approx 90^{\rm o}$, even without exactly knowing
the correlation between activity, rotation and spectral type. A large sample of stars is 
needed in order to perform a previous study of the activity-$v\sin i$ relations 
for the different spectral types. However, some complications arise when trying 
to compile $\log(R'_{HK})$ and $v\sin i$ data, since the current measurements 
are too scarce and imprecise. A solution to this is the simulation of a large 
sample of stars (Sect. \ref{simul}).

The basic assumption in a selection of highly inclined stars for a targeted
transit search lies on the alignment between the stellar rotation and the
planet's orbital spin axis. Although from conservation of angular
momentum we would expect the planet to orbit close to the stellar equator
plane, recent spectroscopic observations during exoplanet transits have
revealed significant spin-orbit missalignments for 10 of 26 Hot Jupiters
\citep{2010A&A...524A..25T} through the Rossiter-McLaughlin effect
\citep{1924ApJ....60...15R,1924ApJ....60...22M}. These planets are thought to
have formed far out from the star and migrated inwards. In this
process, planet-planet scattering and Kozai oscillations due to additional 
companions could significantly affect the obliquity of the orbit
\citep{2003ApJ...589..605W,1996Sci...274..954R}. In spite of scarce statistics,
spin-orbit missalignments have only been observed in Hot Jupiters, and
assuming that a planet has formed and migrated in a disc, it is expected the
majority to be in aligned orbits. Several multiple transiting systems have
recently been found by Kepler \citep{2011Natur.470...53L,2011arXiv1102.0543L},
giving more weight to the existence of planets with spin-orbit aligment.

It is also worthwhile noting that \cite{2010A&A...512A..77L} reported evidence
that Hot-Jupiters orbiting close to their stars can affect their angular
momentum evolution by interaction with their coronal fields
\citep{2009A&A...505..339L}. This would complicate our approach, as slightly
different rotation rates would be expected for stars with giant planets,
preventing us to predict the stellar axis inclination from the same
activity-$v\sin i$ distribution as for general stars. However, this effect has
only been observed for hot giant planets orbiting very close to early-type
stars, and the possible induced rotation rate bias is smaller than the typical
$v\sin i$ precisions. Therefore, no actual complexities are added.

\section{Stellar sample simulation}
\label{simul}

\subsection{Aims \& assumptions}
\label{aims}

The simulation of a large sample of stars containing the basic data for the
subsequent analysis is needed if we desire to study accurately the correlation
between chromospheric activity and projected rotational velocity. The resulting
sample should follow the available activity and $v\sin i$ data from different
existing catalogs, including the effects of observational errors and cosmic
dispersions for the different parameters. A simulation of this kind can be
achieved using evolutionary models and several empirical relations.

Among the output data of the simulation will be the stellar projected inclination
determined from an isotropic distribution of the rotation axes. Different
inclinations cover different areas in the activity-$v\sin i$ diagram, and
whereas the current available observed data are too scarce to properly study
their distribution and trends, the simulation presented here will allow us to
accurately study how are they distributed in the diagram and perform a
selection of observed stars from real data using the method described in
Sect. \ref{stats}. The possible bias resulting from our selected sample will
be discussed in Sect. \ref{discus}. The main idea of the method is not to
determine stellar inclinations of real stars, but to estimate the probability
for each star to have higher inclination than a certain value. This means,
under the assumption of spin-axis alignment, defining a subset of stars for
which a transit search would be most efficient.

Our study is limited to G, K, and M dwarfs ($0.6<(B-V)<1.6$). This is mainly
because we require a convective envelope to assume the connection between the
rotation rate and the level of stellar activity. Therefore, we are constraining 
$B-V$ so that the spectral range covered by our selection strategy contains all 
the stars where the activity has been observed to be scaled by the rotation period. 
This includes up to late M-type stars, so the same analysis can be performed there 
 \citep{2003ApJ...583..451M,2009ApJ...693.1283W,2011ApJ...727...56I}.
However, our method finds some limitations when applied to M-type stars from
the fact that magnetic activity is known to rapidly saturate at a given level
as rotation rate increases. A more extensive discussion on this is
given in Sect. \ref{discus}.

\subsection{Generation of the sample}
\label{gen}

The masses of the simulated stars are generated so that they follow the
distribution of the Present Day Mass Function for the solar neighborhood by
\cite{1979ApJS...41..513M} and limited to within 0.15 $\rm M_{\odot}$ and 1.05
$\rm M_{\odot}$. These limits account for the spectral range we are
interested to cover. The upper limit is set at stars of spectral type G0. For
higher mass stars the depth of the convection is increasingly small and
hence stellar activity may be significantly less intense 
\citep{1982ApJ...258..201G}. On the other hand, we exclude stars later than $\sim$M4 where
 as will be shown in Sects. \ref{stats} and
\ref{data}, activity-rotation pattern leaves little chance to our
selection method considering the precision of the current $v\sin i$ data.

Our sample is restricted to main sequence stars as magnetic activity behavior
in evolved stars is still not well understood. Therefore, the stellar ages are
generated considering the main sequence lifetime, which depends on the stellar
mass and was calculated by fitting the terminal age main sequence points from
the evolutionary tracks of \cite{2004ApJ...612..168P}, generated using the
BaSTI web tool\footnote{http://albione.oa-teramo.inaf.it/}. We obtained the 
following expression for the main sequence lifetime $\tau_{ms}$:
\begin{equation}
\begin{array}{l}
\log\tau_{ms}=9.92-3.85\log M/{\rm M_{\odot}}+2.50(\log M/{\rm M_{\odot}})^{2}-\\
-1.67(\log M/{\rm M_{\odot}})^{3}\\
\end{array}
\label{tms}
\end{equation}
In fact, the upper limit on age only affects G-type stars, as the ages of
lower-mass stars are limited by the range covered by the evolutionary models
used later to compute stellar radius and color index
\citep{2008A&A...482..883M}.
As a result, the age distribution for most of the spectral range is flat,
covering from 0.1 to 12.6 Gyr. The oldest low-mass stars are expected
to be inactive and very slow rotators \citep{2007ApJ...669.1167B},
%as predicted by the evolution described in Appendix \ref{rotage}
and hence they would show very low values of $v\sin i$. As will
be seen later in Sects. \ref{stats} and \ref{data}, such $v\sin i$ values are
likely to be far from the measurement possibilities of current spectrographs
if we want to distinguish and select different ranges of stellar inclinations.
Therefore, the exclusion of old K and M-type stars does not introduce any limit
to the optimization method that we are designing.

The ($B-V$) color indices are derived from mass and age values using stellar
models \citep{2008A&A...482..883M}. As we are simulating samples of stars in the solar
neighborhood, interstellar absorption is negligible and no reddening is considered. The simulated sample is
limited to a specific range in ($B-V$) between 0.6 and 1.6. This is important as
one of our goals is to study the variations in the activity-rotation behaviour
with spectral type and how this can affect the possibility of resolving
different ranges of inclination. The statistics described in Sect. \ref{stats}
will help to set some criteria concerning the validity of our selection method
at different ($B-V$) ranges.

Once ($B-V$) and age is known for a given star, the rotation period can be obtained using
known empirical relations. Around the age of most of open clusters, many
observations converge to follow a $t^{1/2}$ spin-down law. In
\cite{2003ApJ...586..464B} and \cite{2007ApJ...669.1167B}, observations from
several open clusters are used to obtain the rotation rate as a function of
spectral type for F, G and K stars and calibrating the age dependence using the
Sun. The age-rotation relations 
by Engle et al. 2011 (in preparation) were obtained from stars in the spectral range covered by the simulation and which have 
age determinations through different indirect methods, and thus
represent a more suitable approach for our purposes. They follow an
empirical expression of the form:

\begin{equation}
\label{eqrotage}
P_{rot}(\rm days)=P_{0}+a\cdot [Age (\rm Gyr)]^{b}
\end{equation}

%presented in Appendix \ref{rotage} of this work 
\noindent for a specific ($B-V$). However, as will be
commented in Sect. \ref{data}, very similar results are obtained if the 
expression of \cite{2007ApJ...669.1167B} is used instead.

Three independent empirical relations of the form of expression \ref{eqrotage} were 
obtained for G-, K- and M-type dwarfs. The coefficients of the best fit are presented in Table \ref{tab:coefs}
Then, the mean ($B-V$) was calculated for each of the 
three samples. Knowing the ($B-V$) of our simulated stars, their period was obtained 
by interpolating the results given by the three relations.
A Gaussian error was
added to the rotational periods obtained in order to consider a cosmic dispersion
around the age-rotation relations. The standard deviation for this Gaussian
distribution was estimated from the standard deviation of the rotational period data 
provided by Engle et al. 2011 (in preparation),
%presented in Appendix \ref{rotage}
being near 1 day and having a slight dependence on age.

\begin{table}
\caption{Coefficients for the age-rotation relations (expression \ref{eqrotage}) used in the simulation.}
\label{tab:coefs}
\centering
\begin{tabular}{cccc}
\hline\hline
Sp. Type & Parameter & Value & Std. error\\
\hline
G & $P_{0}$ & 2.361 & 2.722 \\
 & $a$ & 6.579 & 3.609 \\
 & $b$ & 0.763 & 0.253 \\
\hline
K & $P_{0}$ & 2.158 & 2.092 \\
 & $a$ & 10.224 & 2.922 \\
 & $b$ & 0.663 & 0.131 \\
\hline
M & $P_{0}$ & -6.624 & 13.107 \\
 & $a$ & 25.797 & 15.333 \\
 & $b$ & 0.711 & 0.268 \\ 
\hline
\hline
\end{tabular}
\end{table}

\begin{figure}
\centering
   \includegraphics[width=\columnwidth]{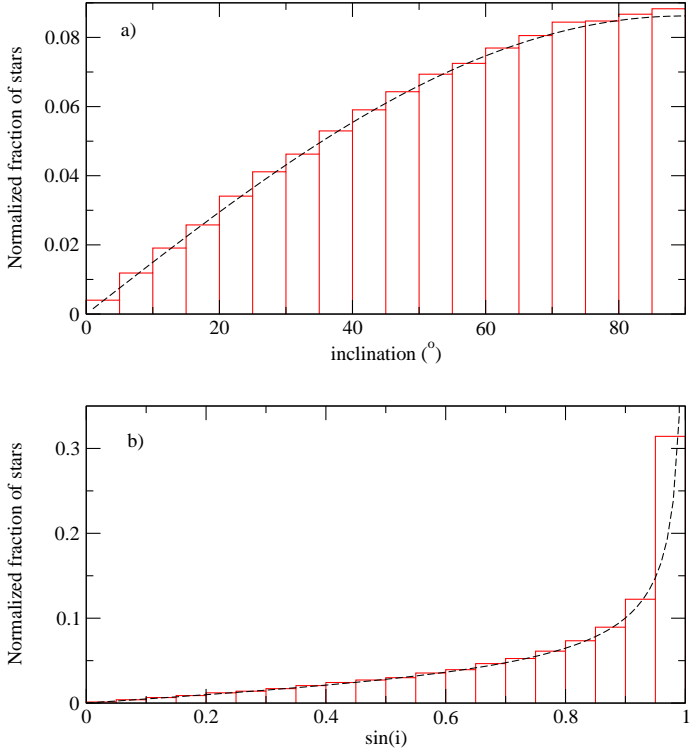}
    \caption{a) Frequency of stars in the simulated sample depending on the
inclination of their rotation axis towards our line of sight, considering an
isotropic distribution resulting from the simulation. b) Frequency of $\sin i$
values in the same distribution. The behavior of the sine function, weighed
toward $\sin i=1$, makes difficult to select highly inclined stars. The analytic 
functions that describe the distributions are also plotted in both graphs (dashed lines).}
     \label{fig:hist_inc}
\end{figure}

The stellar radius is obtained from linear interpolation of the evolutionary
models of \cite{2008A&A...482..883M} and then the equatorial rotation velocity
can be calculated from the rotational period. An isotropic distribution of
rotation axis orientations is assumed to generate the inclination of each
simulated star, thus giving about $9\%$ of the sample with $i>85^{\circ}$ and
$17\%$ with $i>80^{\circ}$ (see Fig. \ref{fig:hist_inc}). An additional Gaussian
dispersion component is added to the resulting $v\sin i$ values to include the
measurement uncertainty. High-resolution spectroscopy and cross-correlation
techniques are currently yielding $v\sin i$ measurements with errors lower than
$\sim0.5$ km s$^{-1}$ \citep{1999A&AS..139..433D,2003csss...12..823G}.

Regarding stellar activity, several empirical relations exist that correlate 
Ca~{\sc ii} H and K emission with rotational period and spectral type. Due to the
lack of sufficient data for an accurate study, the description of the stellar
chromospheric activity in terms of the stellar dynamo has been difficult
\citep{1996ApJ...466..384D,2001MNRAS.326..877M}. \cite{1984ApJ...279..763N}
reported rotation periods for a sample of main sequence stars and demonstrated
that chromospheric flux scales with the Rossby number, $R{\rm o}=P_{\rm
rot}/\tau_{c}$, where $P_{\rm rot}$ is the rotation period and $\tau_{c}$ is
the convective overturn time near the bottom of the convection zone, which is
an empirical function of the spectral type. This gives a much better
correlation than activity-period, but obtaining $R{\rm o}$ is complex due to the
difficulties in measuring stellar rotation rate and the estimation of the
turnover time based on stellar interior models \citep{1996ApJ...457..340K}.
\cite{1984ApJ...279..763N} give an empirical calibration for the turnover time
in terms of ($B-V$), considering an intermediate value for the ratio of mixing
length to scale heigth, $\alpha =1.9$. The fit is given by the polynomial
expression:
\begin{equation}
\label{tau}
$$
\log\tau_{c}=\left\{ \begin{array}{rl}
1.362-0.166x+0.025x^{2}-5.323x^{3}, &\mbox{ x$>$0} \\
1.362-0.14x,&\mbox{ x$<$0}
\end{array} \right.
$$
\end{equation}
\noindent where $x=1-(B-V)$ and $\tau_{c}$  is the turnover time in days. This is used in our simulation to compute the
convective turnover times, and then the Rossby numbers for each star. However,
as \cite{1980LNP...114...19G} and \cite{1984ApJ...279..763N} show, different
values of $\alpha$ make stars deviate from Eq. (\ref{tau}), although the
best correlation is found for $\alpha =1.9$ considering mixing-length theory
models. In our calculated convective turnover times, a Gaussian dispersion 
with $\sigma _{\log \tau_{c}} =0.03$ dex is added to the $\log \tau_{c}$ values 
to take this dispersion into account. \cite{2008ApJ...687.1264M}
give a relation between the chromospheric index $\log(R'_{HK})$ and the Rossby
number from a larger sample, which we use to obtain the activity for each of
our simulated stars. The calculation of the final $\log(R'_{HK})$ also accounts
for a Gaussian dispersion as the data in \cite{2008ApJ...687.1264M} show. The 
amplitude of this Gaussian dispersion, $\sigma _{\log(R'_{HK})}$, is scaled 
with $R{\rm o}$ being always near 0.04 dex.

\subsection{Statistics on the simulated samples}
\label{stats}

All the parameters from the simulated stars are generated from empirical
relations, so the resulting sample should properly account for the trends that
real observed stars show in the activity-$v\sin i$ diagram. The large error
bars in many of the available $v\sin i$ measurements (being some of them only
upper limits) and the selection biases in the observed catalogs make it
difficult to compare our simulated sample with observations. Therefore, 
a careful selection of observational catalogs will be
necessary to correctly check the agreement with the simulated data. The
Gaussian dispersions introduced to the different empirical relations at several
steps of the simulation and the observational uncertainty added to the obtained
$v\sin i$ values, as described in Sect. \ref{aims}, are critical to obtain a
simulated sample which properly fits real data.

\begin{figure}
\centering
   \includegraphics[width=\columnwidth]{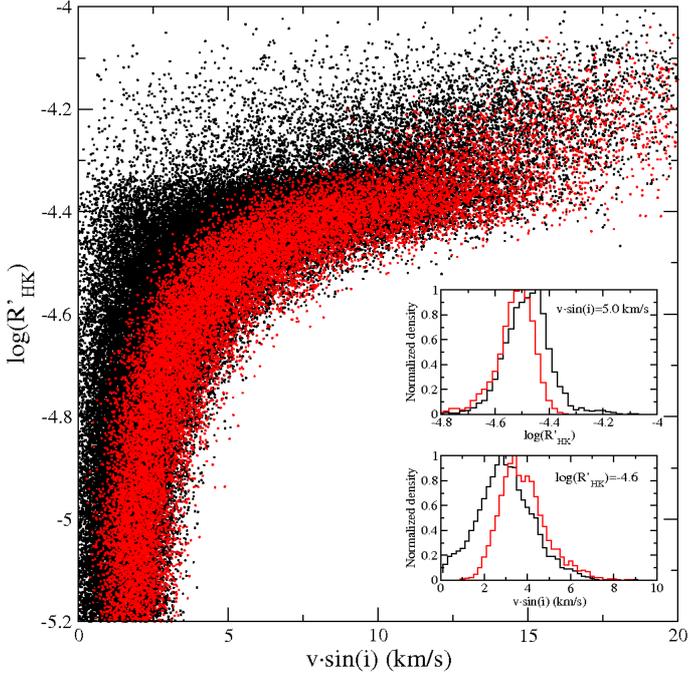}
    \caption{A sample of 100,000 solar-type stars ($0.6<(B-V)<0.7$) simulated
with the procedure described in Sect. \ref{aims}. Random errors with a Gaussian
distribution ($\sigma =0.5$ km s$^{-1}$) were added to the $v\sin i$ values.
Stars with projected axis inclinations above $80^{\circ}$ are represented in
red and trace the envelope region at the right of the distribution. For a
constant $\log(R'_{HK})$, stars are expected to have a very similar equatorial
velocity in terms of the rotation and activity evolution assumptions, and so
different $v\sin i$ are mainly due to different axis projections. The 
histograms in the two overlayed plots show the density distribution of the 
simulated samples at two cuts made at constant $\log(R'_{HK})=-4.8$ and 
$v\sin i=7.5 \rm{km/s}$ respectively.}
     \label{fig:act_inc}
\end{figure}

\begin{figure*}
\centering
    \resizebox{\hsize}{!}{\includegraphics{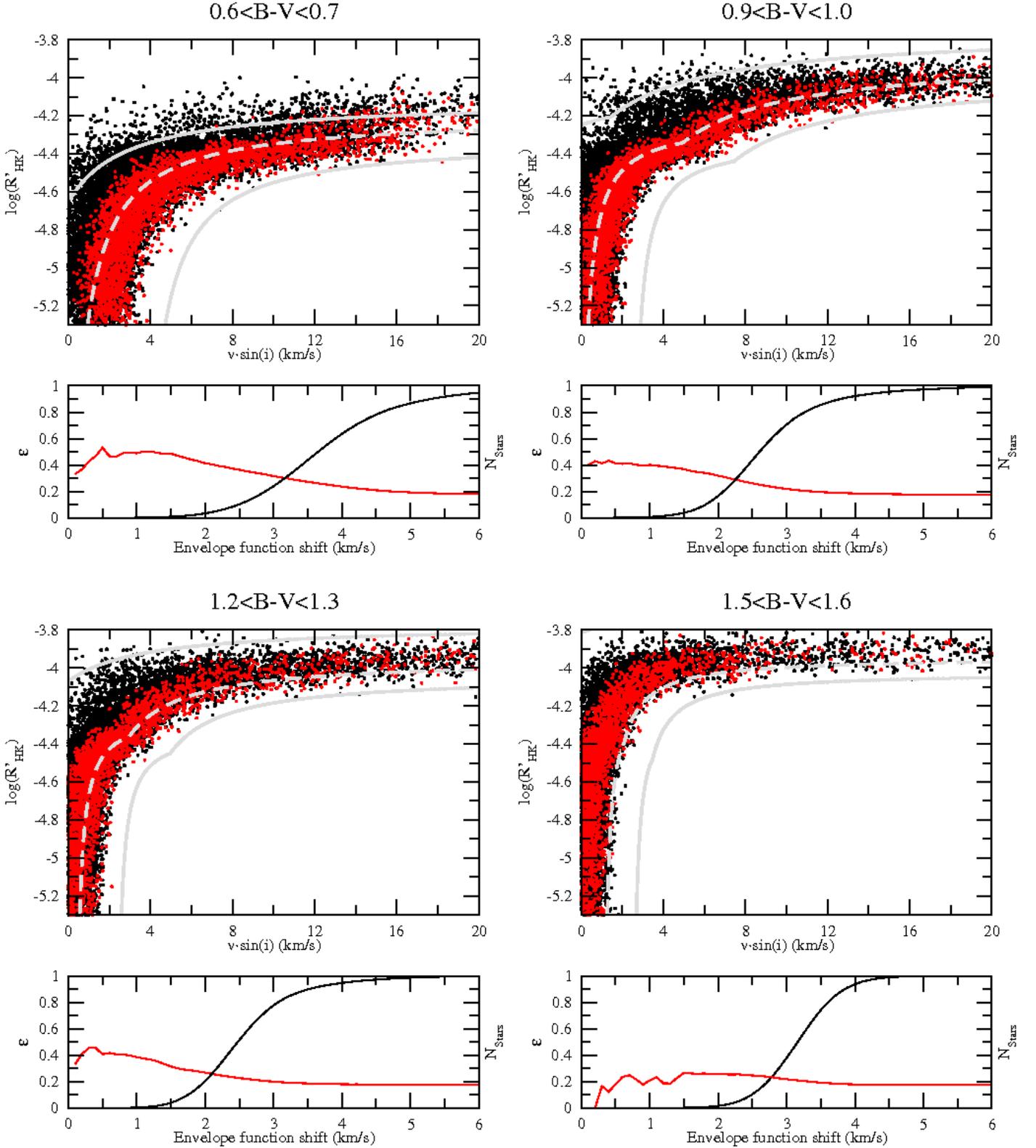}}
    \caption{Four simulated samples limited in colour index as indicated. Stars
with $i>80^{\circ}$ are plotted in red. The grey envelope function, 
calculated as described in Sect. \ref{stats}, is shifted
from the right part of the distribution to the left considering a slope so that
it scans the high inclination region (see text). At each step the efficiency
(Eq. \ref{eff}, with $\alpha=80^{\circ}$) is calculated for the subsample below
the envelope function. The red line in the bottom diagrams shows the evolution
of $\epsilon$ as the region limited by the function is expanded to the left,
and the black line indicates the total fraction of stars in the region
($N_{Stars}\sim 1$). As expected, $\epsilon$ tends to $\sim 0.17$ for
$N_{Stars}\sim 1$. The dotted grey line in the upper diagrams indicates the
region where $\epsilon=0.26$, thus the efficiency is increased by 50\% with respect to a
non-preselected sample.} 
     \label{fig:evol_stats}
\end{figure*}

As expected, for all the spectral ranges our simulations show a clear increase
in the mean projected rotation velocity for higher chromospheric activity (see
Fig. \ref{fig:act_inc}). Also, stars with inclinations close to $90^{\circ}$
are placed at the right side of the distribution, thus defining the
 envelope region of interest, clearly distinguishable from stars with other
orientations. The contamination comes from the cosmic dispersions and
observational uncertainties. As it is evident from Fig. \ref{fig:act_inc}, the
distribution in activity-$v\sin i$ becomes more densely concentrated for later
spectral types, because inactive stars tend to be very slow rotators,
thus requiring very precise $v\sin i$ measurements in order to apply our selection method. The discontinuity in the
expression used to compute $\log(R'_{HK})$ from the Rossby numbers
\citep{2008ApJ...687.1264M} for the simulated stars causes a small bump in the
activity-$v\sin i$ distribution near $\log(R'_{HK})=-4.35$, which is more
evident for later spectral types (see Figs. \ref{fig:act_inc} and 
\ref{fig:evol_stats}). Although this could be avoided by using a single
expression for the calculation, like the one obtained by
\cite{1984ApJ...279..763N}, it would be at the expense of losing precision in
the general shape of the distribution.

With sufficient stars in the activity-$v\sin i$ diagram and the possibility of
generating a sample for any spectral type, we can define an efficiency
parameter for each location in the diagram. This is defined as
the probability for a star in that location to have an inclination angle above
a certain angle, and can be calculated by defining a small region in the
diagram around the star and then computing the ratio:
\begin{equation}
\label{eff}
\epsilon_{\alpha}=\frac{N_{i>\alpha}}{N_{T}}
\end{equation}
\noindent where $N_{i>\alpha}$ is the number of simulated stars in the region
with axis inclinations above a given $\alpha$ angle and $N_{T}$ is the total
number of stars in the same region. For a specific ($B-V$)-limited sample, and
given fixed cosmic dispersions and Gaussian errors for the observables, the
efficiency $\epsilon$ is a property of each location in the activity-$v\sin i$
diagram, and therefore regions of interest for transit searches can be studied.
 To quantify the increase effectiveness of the present methodology, we 
assume $\alpha=80^{\circ}$ as most of the known transiting planets are found 
above this inclination angle. In a global sample with an isotropic distribution of
rotation axes, we would expect 17.3\% of the stars to have $i>80^{\circ}$, so
that $\epsilon_{T}=0.173$. We may define the normalized efficiency as:
\begin{equation}
\label{neff}
P=\frac{\epsilon_{80^{\rm o}}}{\epsilon_{T}}=\frac{1}{\epsilon_{T}}\frac{N_{i>80^{\rm o}}}{N_{T}}
\end{equation}
\noindent so that $P=1$ for a non-selected sample and $P>1$ for
selected subsamples with increased high inclination probability.

While the efficiency for a non-selected sample of stars would be $\epsilon\approx
0.173$, this increases as we move to regions at the right side of the
distribution in the activity-$v\sin i$ diagram, so increasing $P$. For simulated samples limited
in spectral type, an envelope polynomial function of the form $y=a_{1}+a_{2}/x$
was calculated by fitting the subset of stars with $i>85^{\circ}$, splitting
the function into two in order to account for the inactive and the active part of
the distribution. Successive shifts were aplied to the envelope function while
the efficiency was calculated considering all the stars lying in the region
below it (see Fig. \ref{fig:evol_stats}). One can study the efficiency changes
and test the applicability of the selection method for different specral types
by scanning the activity-$v\sin i$ diagram with the envelope function that
traces equal-inclination regions and performing such statistics.

Four samples of different spectral types are represented in Fig.
\ref{fig:evol_stats}. In all cases, highly inclined stars trace the envelope
region of the distribution, but considering the observational and cosmic
dispersions, very inactive or active stars (at the saturation zone) for later
spectral types become indistinguishable in the activity-$v\sin i$ diagram in
terms of stellar inclination. The empirical errors of the activity calculation
(see Sect. \ref{aims}) and a $\sigma=0.5$~km~s$^{-1}$ for $v\sin i$ were
considered in all cases. The range of ($B-V$) is also critical, as rotation rate
can change significantly with stellar mass in the Main Sequence. Figure
\ref{fig:evol_stats} highlights this and also shows how efficiency decays as
the width of the envelope region increases towards the upper-left part of the
distribution and more stars are included in the statistics.

The envelope function previously obtained is first shifted a certain amount to
increasing $v\sin i$ until all active stars lie at lower $v\sin i$. Then, it is
shifted back to decreasing $v\sin i$ at very small steps, considering a slope
so that it scans regions of equal inclination. The two lines define the boundaries of the region
of interest where the statistics is calculated. As expected, the efficiency
decreases as more stars from the part of the distribution with lower $v\sin i$
are included in the region. For the earlier spectral types, a region containing the 
15\% of the entire sample yields
$P\approx 2.0$, which represents doubling the probability of
$i>80^{\circ}$ with respect to a non-preselected sample. M-dwarfs show a less discriminating 
distribution in the diagram and thus the efficiency only reaches $\sim1.5$ (50\% increase than a
non-preselected sample) for a small region containing 1.5\% of the total
simulated sample.

A similar approach to the previous statistics permits the calculation of the
efficiency at a specific location in the diagram. For a real observed star
that is to say the probability for this star to have an inclination angle 
above 80$^{\circ}$, and thus evaluate its suitability as a candidate for a
transit search. Instead of performing the calculation of the efficiency ratio
(Eq. \ref{eff}) in a wide region as before, a small region around the point of
interest can be defined. For real data, we determine this region from the
observational uncertainties in $v\sin i$ and $\log (R'_{HK})$. Also,
observational errors in ($B-V$) for the star are taken into account for limiting
the spectral range of the simulated sample used to calculate $P$. In this case,
the Gaussian errors for the observables are not introduced when generating the
simulated data points, as they are already taken into account when defining the
box where the efficiency is calculated.

\section{Data selection and analysis}
\label{data}

The main goal of the generation of a simulated sample of stars is to provide
the basis to perform the statistics described in Sect.  \ref{stats} to real
data. Currently, several catalogs exist that compile high-resolution
spectroscopy measurements of both Ca~{\sc ii} H and K fluxes and projected rotation
velocities. However, special care has to be taken when selecting the data and
cross-matching catalogs, as different authors use different spectral
resolutions and reduction techniques, and this results in a diversity of
precisions, detection limits and possible selection biases.

A first test to the high-inclination selection method is to perform
the efficiency statistics on the stars which already have known transiting
planets. Although several of them have been observed to be spin-orbit
missaligned \citep{2010A&A...524A..25T}, the majority of the stars with
transiting planets are expected to have inclinations near $90^{\circ}$, and
thus we should obtain a high efficiency rate when calculating the statistics
with the properly generated star sample for each object. It is important to
notice that while the Rossiter-McLaughlin effect provides a measurement of the
spin-orbit angle on the plane of the sky, there is still a component to be
determined, and this is related to the stellar inclination and the planet orbit
angle towards the line of sight. Therefore, the information given by the Rossiter-McLaughlin measurements is
complementary to the one our statistical method can provide, since both angles are independent.

Also relevant are the results from \cite{2011arxiv1103.5332} showing that no
significant correlations exist between chromospheric activity indicator $\log
(R'_{HK})$ and planet presence or parameters. This is important if we want to
apply the described activity-$v\sin i$ statistics to these stars, using the
same empirical approach used in Sect. \ref{simul} for the observed stars.

Lists of $v\sin i$ measurements for stars with planets can be found in
\cite{2010ApJ...719..602S} and \cite{2010MNRAS.403.1368G}, while
\cite{2010ApJ...720.1569K} compile both $\log(R'_{HK})$ and $v\sin i$ data from
several authors. The best quality measurements of the subset of G, K and M-type
stars were selected and are presented in Table \ref{tab:trans}. When available,
the error bars of the listed parameters were used to constrain the generated
sample and the box where the statistics is calculated. Otherwise, a typical box
size of 0.1 dex in $\log(R'_{HK})$ and 0.5 km s$^{-1}$ in $v\sin i$ was
adopted. For each star, a simulated sample of $10^6$ stars was generated,
covering a range around its $(B-V)$ value (0.04 magnitudes when no errors are
available for photometry), and the efficiency was calculated with the
simulated stars inside the defined box in the activity-$v\sin i$ diagram.

\begin{table}
\caption{List of G-K-M type exoplanet parent stars with available data.
Activity-$v\sin i$ distributions of simulated samples where used to calculate
the normalized efficiency parameter $P$ (see text). This gives the probability
of each star to have a rotation axis inclination above $80^{\circ}$, divided by
the same probability considering a sample with an isotropic distribution
of rotation axes.}
\label{tab:trans}
\centering
\begin{tabular}{cccccc}
\hline\hline
Star &  ($B-V$) & $\log(R'_{HK})$ & $v\sin i$ (km/s) & $P$ & Ref. \\
\hline
HD 17156 & 0.64 & -5.022 & 5.0$\pm$0.8 & - & 1,3 \\
HD 80606* & 0.77 & -5.10$\pm$0.04 & 2.0$\pm$0.4 & - & 3,4 \\
HD 149026 & 0.61 & -5.030 & 6.0$\pm$0.5 & - & 1,5 \\
HD 189733 & 0.93 & -4.50$\pm$0.05 & 4.5$\pm$0.5 & 2.38 & 3,4 \\
GJ 436 & 1.52 & -5.23$\pm$0.02 & 1.0$\pm$0.9 & 1.68 & 4,6 \\
TrES-1* & 0.78 & -4.738 & 1.1$\pm$0.3 & 0.80 & 1,7 \\
TrES-2 & 0.62 & -4.949 & 2.0$\pm$1.0 & 0.97 & 1,2 \\
TrES-3 & 0.71 & -4.549 & 1.5$\pm$1.0 & 0.09 & 1,8 \\
CoRoT-2 & 0.69 & -4.331 & 11.8$\pm$0.5 & 2.09 & 1,9 \\
CoRoT-7 & 0.80 & -4.802 & 1.3$\pm$0.4 & 1.04 & 1,10 \\
Kepler-4 & 0.62 & -4.936 & 2.2$\pm$1.0 & 1.06 & 1,11 \\
Kepler-6 & 0.68 & -5.005 & 3.0$\pm$1.0 & 3.06 & 1,12 \\
WASP-2* & 0.84 & -4.84$\pm$0.10 & 1.6$\pm$0.7 & 1.61 & 13 \\
WASP-4 & 0.74 & -4.50$\pm$0.06 & 2.0$\pm$1.0 & 0.37 & 1,14 \\
WASP-5 & 0.66 & -4.72$\pm$0.07 & 3.5$\pm$1.0 & 1.70 & 13 \\
WASP-13 & 0.60 & -5.263 & 2.5$\pm$2.5 & 1.00 & 1,15 \\
WASP-19 & 0.70 & -4.660 & 4.0$\pm$2.0 & 1.56 & 1,16 \\
XO-1 & 0.69 & -4.958 & 1.11$\pm$0.67 & 0.80 & 1,17 \\
XO-2 & 0.82 & -4.988 & 1.3$\pm$0.3 & 2.78 & 1,18 \\
HAT-P-3 & 0.87 & -4.904 &0.5$\pm$0.5 & 0.67 & 1,19 \\
HAT-P-10 & 1.01 & -4.823 & 0.5$\pm$0.2 & 0.02 & 1,20 \\
HAT-P-11 & 1.02 & -4.584 & 1.5$\pm$1.5 & 0.99 & 21 \\
HAT-P-12 & 1.13 & -5.104 & 0.5$\pm$0.4 & 1.02 & 1,22 \\
HAT-P-13 & 0.73 & -5.138 & 2.9$\pm$1.0 & - & 1,23 \\
HAT-P-15 & 0.71 & -4.977 & 2.0$\pm$0.5 & 2.55 & 1,24 \\
%HAT-P-18 & 0.93 &  & 0.5$\pm$0.5 & & \\
%HAT-P-19 & 0.74 &  & 0.7$\pm$0.5 & & \\
\hline
\hline
\end{tabular}
\begin{list}{}{}
\item[$^{\mathrm{*}}$] Spin-orbit misalignment detected through
Rossiter-McLaughlin effect.
\item References: (1) \cite{2010ApJ...720.1569K}; (2) \cite{2010ApJ...719..602S}; (3)
\cite{2010MNRAS.403.1368G}; (4) \cite{2004ApJS..152..261W}; (5)
\cite{2005ApJ...633..465S}; (6) \cite{2003csss...12..823G}; (7)
\cite{2007PASJ...59..763N}; (8) \cite{2009ApJ...691.1145S}; (9)
\cite{2008A&A...482L..25B}; (10) \cite{2010A&A...519A..51B}; (11)
\cite{2010ApJ...713L.126B}; (12) \cite{2010ApJ...713L.136D}; (13)
\cite{2010A&A...524A..25T}; (14) \cite{2009A&A...496..259G}; (15)
\cite{2010A&A...512A..77L}; (16) \cite{2010ApJ...708..224H}; (17)
\cite{2006ApJ...648.1228M}; (18) \cite{2007ApJ...671.2115B}; (19)
\cite{2007ApJ...666L.121T}; (20) \cite{2009ApJ...696.1950B}; (21)
\cite{2010ApJ...710.1724B}; (22) \cite{2009ApJ...706..785H}; (23)
\cite{2009ApJ...707..446B}; (24) \cite{2010ApJ...724..866K}.
\end{list}
\end{table}

The efficiency values presented in Table \ref{tab:trans} were calculated using
Eq. (\ref{eff}) with $\alpha=80^{\circ}$, and then normalizing by the factor
0.173 as explained in Sect. \ref{simul}. The normalized efficiency ($P$) is 
the factor by which the probability of
having $i>80^{\circ}$ is increased (compared with a non-preselected sample of
stars). Because of the uncertainties considered for the box size where the
statistics was calculated, very few transiting host stars show a normalized
probability $P>2$.  However, the mean probability of $i>80^{\circ}$ for this
sample is 1.41, about 40\% higher than for a general sample (expected to give 
$\overline{P}\approx 1.00$). From the 21 stars sample
where the statistics was performed, 13 show $P>1.00$ and 9 show $P>1.50$. Our 
strategy would select these objects, and hence a number of the transiting 
planets in the list could have been found in the subsequent targeted search.

\begin{figure}
\centering
   \includegraphics[width=\columnwidth]{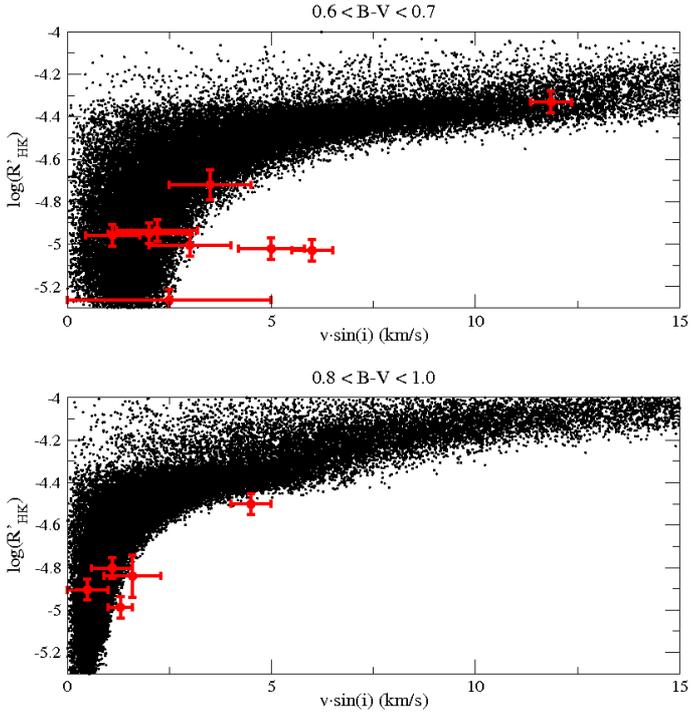}
    \caption{Two subsets of transiting planet host stars (red symbols): solar
type (top diagram) and early K-type (bottom diagram). Individual color indices
and errors are used to constrain the ($B-V$) range of the simulated sample for
each star and error bars were used to define the box to calculate the
statistics (see text). Note that real data points are shown for simulated
samples of a wider spectral range for convenience.}
     \label{fig:trans}
\end{figure}

As can be seen in
Fig. \ref{fig:trans}, where two subsets of different spectral types are shown,
most of the stars lie near the right side region of the simulated samples. Four
of them are even located to the right of the entire distribution, and thus no
statistics could be performed around them. These cases should be investigated
in more detail and are probably due to underestimated $v\sin i$ error bars.
Alternatively, they could be explained by anomalously low values of $\log
(R'_{HK})$, corresponding to a deep minimum of the activity cicle, but 
this is deemed quite improbable.

The next step is the application of the formalism to catalogs of $\log
(R'_{HK})$ and $v\sin i$ measurements with the aim of selecting a sample with
higher probability of high spin axis inclinations. In this case, a result of
$P=2.00$ or even $P=1.50$ can be considered interesting. For example, the
preselection of a sample of stars having $P_{mean}=1.50$ would represent a 50\%
increase on the ratio of highly inclined stars, a thus a 50\% increase on the
efficiency of a planet transit search assuming a spin-orbit alignment.

As it is evident from Fig. \ref{fig:evol_stats}, the performance of the
selection method that we are developing is more efficient for stars between certain
activity thresholds. It is difficult to constrain the inclination
angle for fast rotators at the saturated activity regime, as a wide range of
$v\sin i$ values show very similar activity levels at this early stage of the stellar
evolution. On the other hand, inactive stars for most spectral types present rotation
rates that are too low to resolve different stellar inclinations given the
dispersion of the empirical relationships and the uncertainties of
observational quantities. With regards to these, the activity-$v\sin i$ relation
makes the statistics difficult or unreliable for higher ($B-V$) indices (see
Sect. \ref{stats}), and these should be excluded from the selection process
in order to avoid false positives. A $\log (R'_{HK})$ threshold depending on
($B-V$) was calculated by considering a $v\sin i\geq 1$ km s$^{-1}$ limit and
applying the envelope functions previously used for studying the distribution
(see Sect. \ref{simul}). These functions trace the right edge of the
distribution for each ($B-V$) interval, and thus their cut at $v\sin i=1$ km
s$^{-1}$ can be used to define the inactive limit. This was
done for each 0.1 mag interval $(0.6<(B-V)<1.6)$ and finally a simple polynomic
function was fitted, obtaining $\log (R'_{HK})_{\rm lim}=-3.45-1.36/(B-V)$.
Stars below this inactive limit are not considered for the analysis and
selection process.

\section{A catalog for transit surveys}
\label{catalog}

Ca~{\sc ii} H and K flux measurements of 1296 stars made at Mount Wilson
Observatory were published by \cite{1991ApJS...76..383D}. We converted the $S$
fluxes, together with their uncertainties, to the more standard activity index
$\log (R'_{HK})$ with the method described by \cite{1984ApJ...279..763N}. The
catalog was cross-matched with the list of about 39000 $v\sin i$ measurements
compiled by \cite{2003csss...12..823G}, rejecting those that are upper limits
or have large uncertainties. We further cross-matched the resulting catalog
with the photometry from the 2.5-ASCC \citep{2009yCat.1280....0K} to obtain
color indices. Finally, Main Sequence G-, K- and M-type stars were selected
resulting in 189 objects. Additionally, \cite{2011arXiv1103.0584J} present
chromospheric activity and rotational velocities for more than 850 solar-type
and subgiant stars. Hipparcos distances from \cite{2007ASSL..350.....V} were
used to reject evolved or subgiant objects, as described by
\cite{2011arXiv1103.0584J}, and color indices were obtained from
\cite{2009yCat.1280....0K} resulting in 509 Main Sequence stars in the
spectral range of the analysis. The catalogs with chromospheric activity data
of \cite{1996AJ....111..439H}, \cite{2004ApJS..152..261W} and
\cite{2006AJ....132..161G} were also cross-matched with the rotational
velocities by \cite{2003csss...12..823G} and the color indices from the
2.5-ASCC. Finally, \cite{2010A&A...514A..97L} present spectroscopic data,
including rotational velocities and Ca~{\sc ii} H and K flux for 57 main
sequence stars in the $0.6<(B-V)<1.6$ range.

The complete list of catalogs of G-, K- and M-type ($0.6<(B-V)<1.6$) Main
Sequence stars compiled here is presented in Table \ref{tab:res}, together with
the total number of stars and the number of those found to have a high
inclination normalized efficiency over 1.5, 2.0 and 2.5. All the results,
including the efficiency calculated for each individual star from the input
catalogs, are shown in Table 3 (available online).

\begin{table}
\caption{List of analysed samples of stars with $\log (R'_{HK})$, $v\sin i$ and
(B-V) data. Only G-, K- and M-type stars in the main sequence are included.}
\label{tab:res}
\centering
\begin{tabular}{ccccc}
\hline
Data source &  N. of & \multicolumn{3}{c}{Number of stars with} \\
 & stars & $P>1.5$ & $P>2.0$ & $P>2.5$ \\
\hline
Dc91, Gb03 \& Kh09 & 189 & 37 & 29 & 16 \\
Jk10 \& Kh09 & 509 & 209 & 124 & 66 \\
Gy06, Gb03 \& Kh09 & 128 & 37 & 28 & 13 \\
Wr04, Gb03 \& Kh09 & 239 & 64 & 44 & 32 \\
Hr96, Gb03 \& Kh09 & 99 & 24 & 15 & 9 \\
Lp10 \& Kh09 & 57 & 9 & 4 & 2 \\
Total & 1221 & 380 & 244 & 138 \\
\hline
\end{tabular}
\begin{list}{}{}
\item References: Dc91 \citep{1991ApJS...76..383D}, Gb03 \citep{2003csss...12..823G},
Gy06 \citep{2006AJ....132..161G}, Hr96 \citep{1996AJ....111..439H}, Jk10
\citep{2011arXiv1103.0584J}, Kh09 \citep{2009yCat.1280....0K}, Lp10
\citep{2010A&A...514A..97L}, Wr04 \citep{2004ApJS..152..261W}.
\end{list}
\end{table}

From the 1221 stars analysed, a subsample of 380 have
a high inclination probability increased by 50\% ($P>1.5$) or more. This
subsample contains the stars where a planet transit search would be most
efficient. The fact that different source catalogs result on a different ratio
of stars with $P>1.5$ is caused by the existence of biases towards more
active or inactive stars in the different catalogs and the fact that our
selection method is more sensitive to the mid and active regime of the
activity-$v\sin i$ diagram (see Sect. \ref{stats}). Tighter sample selections
would produce a higher rate of transit findings while observing fewer stars.
This is a trade-off worth considering. For example, 244 (20\%) of the input
stars show $P>2.0$ and 138 (11\%) give $P>2.5$, which means more than 0.43
probability of being equator-on ($i>80^{\circ}$). Therefore, assuming spin-orbit 
alignment, at least 43\% of the stars that harbor planets in this subset are expected to show transits. 
In general, for stars with spin-orbit aligned planets with $R_{star}/a<\cos (80^{\rm o})$ 
and selected with $P>P'$, we can predict that at least $P'\cos (80^{\rm o})$ of them will show transits.

\section{Discussion}
\label{discus}

As shown in Sect. \ref{catalog}, our strategy for targeted transiting planet
searches results in a reduction of the initial stellar sample and an increase
in the probability of finding transits. It is important to stress that we are
not measuring stellar inclinations directly, but just putting constraints via
the estimation of the statistical parameter $\epsilon$ (or $P$). This gives the
probability for each star to have a rotation axis inclination above
$80^{\circ}$, and thus to be oriented nearly equator-on. The probability can
be calculated for every star with $\log (R'_{HK})$ and $v\sin i$ measurements.
A preselection of stars with a high value of $P$ is expected to provide a
considerably higher rate of transiting planets than a non-preselected sample.
Obviously, some stars with transiting planets may not be selected during the 
process and some targets with $i<80^{\circ}$ will inevitably be
included in the selection, but the ratio of stars with $i>80^{\circ}$ will be
always higher (and even more so as we increase the value of $P$) in the
selected sample than in the unselected one, which is the main aim of our
approach.

Several aspects should be taken into account to discuss the credibility of the
resulting probabilities, and thus the validity of the subsequent selected
sample to serve as the input catalog for high-efficiency transit searches.

Firstly, the method for selecting high inclination stars is based on performing
statistics on simulated samples, and thus depends on the accuracy of the empirical 
relations, distributions and dispersions used in the simulation. As we describe in 
detail in Sect. \ref{simul}, these relations were obtained from observed data of stars in the solar
neighborhood that show certain correlations and dispersions. Both the
expressions that describe these correlations and the Gaussian dispersions that
best reproduce the observed data were implemented in the simulation of the
stellar sample, and hence the results are simulated samples of stars that
correctly reproduce the parameters observed in the solar neighborhood. On the other
hand, a better description on both the chromospheric activity and the rotation
rate dependence on mass and age, based on more accurate data, 
could help to better correlate the simulated samples with the observations in
the activity-$v\sin i$ diagrams.

Secondly, the
precision of the measurements of Ca~{\sc ii} H and K flux and $v\sin i$ for
each analysed sample of real stars is important and should be considered, as
very large error bars would make the statistics uncertain and useless.
Some of the currently published measurements of $\log (R'_{HK})$ and $v\sin i$
are quite imprecise or do not have error determinations.
Ca~{\sc ii} H and K flux usually presents variability for active stars, so
several measurements taken at different epochs are needed to determine the
average chromospheric activity and its uncertainty. In this work we considered 
all the objects with available $\log (R'_{HK})$ measurements, applying an error 
box of a mean size of 0.1 dex for the data with no published uncertainties. For 
further study and more accurate results, stars showing chromospheric activity 
variations or having a single $\log (R'_{HK})$ measurement may not be considered 
or should be analysed separately. In the case of $v\sin i$ measurements, these
require high-resolution spectroscopy and a very thorough analysis. Many of the
published values are just upper limits, have large error bars or 
no associated uncertainties. Since the rotation velocities are critical at the
selection process, only data with the best precision was used in
the analysis and results.

Finally, the introduction of some biases and selection effects is evident in our approach, and we should
analyse whether this could influence the planet detection and characteristics.
As described in detail in Sect. \ref{stats}, the statistics provide a better
discriminating power for different values of stellar inclinations at the top
part of the activity-$v\sin i$ diagram, and this prompted us not to consider
the most inactive stars in order to avoid a high level of contamination (see Sect.
\ref{data}). Therefore, our method is much more sensitive for active stars and
the resulting selection
will be biased towards this part of the sample. This means that we are
rejecting part of the slow rotators of the sample, and hence the older stars.
This bias will only have some noticeable effect on late-type stars ($(B-V)>1.0$).
For example, M-type stars older than $\sim 1$ Gyr are expected to have $\log
(R'_{HK})<-4.5$, and the activity-$v\sin i$ distribution for M stars gives no
chance to constrain inclinations in the range below this limit. 

It is important to note that the active/young range of late type stars is the
most unexplored in exoplanet searches. Radial velocity surveys are forced to
reject active stars that tend to show high rotational velocities and radial
velocity jitter. Also, transit photometric surveys (especially ground-based)
are likely to be inefficient for this kind of stars, due to the
time-varying photometric modulations caused by starspots. On the other hand,
selected equator-on stars resulting from our method will have estimated
rotation periods, and hence they are suitable stars for targeted searches where
both the signal of the spot modulation and the possible transits could be
detected and analysed. Therefore, a targeted search based on selected bright
active stars expected to have $i\sim 90^{\circ}$ would be complementary to most
exoplanet searches currently ongoing. The observables required for our method, 
mainly resulting from high resolution 
spectroscopy, require specific equipments and can be time intensive when 
considering large amounts of stars. On the other hand, a single measurement is 
needed for each star, while a photometric monitoring or radial velocity search 
requires long time-series for each of the objects. Moreover, the same data from 
spectroscopic surveys that are required by our methodology can be useful for many 
other purposes. 

Although there is a considerable amount of $\log (R'_{HK})$ and $v\sin i$ data
available nowadays (see Sect. \ref{data}), future high resolution spectroscopic
observations with better precision may help to obtain even more selective
samples for possible targeted transit searches. This strategy, besides being
complementary to the currently ongoing radial velocity and transit surveys, can
be more efficient than a global photometric search with no preselection, which
requires multiple photometric measurements of a large amount of stars to result
in a relatively low rate of transit detections. In addition, targeted
observations carrying out time series photometry of multiple objects are
possible today with the increasing number of small robotic observatories.
Moreover, many amateur astronomers are achieving high precision photometry and
have suitable equipments to take part in a project involving observations of
multiple stars for a transit search. The availability and capabilities of
amateur or small telescopes may represent the most appropiate strategy for a
targeted transit search on bright stars.

\section{Conclusions}
\label{conc}

The main idea of our work was to design and carry out a method to select the
best stellar candidates for a transit search from constraints on their rotation
axis inclination. One feasible way to do so with the currently available
data is to make a statistical estimation of the inclinations by studying the
distribution of the stars from different spectral types in the activity-$v\sin
i$ diagram. The need to perform a simulation of stellar properties arised from the
lack of a large database of $\log (R'_{HK})$ and $v\sin i$ measurements with
sufficient quality, and allowed us to accurately study the distribution of
stars with different inclinations in the activity-$v\sin i$ diagram using the
statistics described in Sect. \ref{stats}. Moreover, the successive steps
made to design the simulation chain use the set of empirical relations that best
describe the properties of the stellar sample in the solar neighborhood. This
can also be useful to other fields.
 
%For example, the presented empirical
%relations for the stellar rotation and age (see Appendix \ref{rotage}) expand and give a more
%accurate approach to our knowledge of the angular momentum evolution of G-, K-
%and M-type stars, which can also be applied to stellar age determinations
%(gyrochronology).

Having the possibility to obtain large simulated samples of stars constrained
in ($B-V$), a relatively simple statistics was designed, so that it can be
performed for every object with ($B-V$), $\log (R'_{HK})$ and $v\sin i$ data. As
a result of calculating the normalized efficiency $P$ (see Sects. \ref{simul}
and \ref{data}) to the stars in the currently available catalogs, we proved
that a preselection of about 10\% of the initial samples can be made achieving
a mean efficiency which is 2 to 3 times better. With the assumption of the
existence of spin-orbit aligned planets around all the stars, this means that
an exoplanet transit search with a 3 times higher success rate can be designed. 
In our work, the application of the approach on more than 1200 stars with currently available 
data has resulted on a catalog containing the most suitable sample for a transit search.

In the future, larger catalogs with more precise measurements of chromospheric
activity and $v \sin i$ will help to derive more accurate relations for the
simulation of the stellar samples, and also to obtain more reliable results for
the selection of highly inclined stars. On the other hand, an observing
strategy considering a targeted exoplanet transit search should be designed
taking advantadge of the incresing availability of small robotic observatories 
and also photometric monitoring nano-satellites that may be launched in the near 
future and that could take profit from the pre-selected samples of stars resulting 
from the presented method.

\begin{acknowledgements} This work was supported by the /MICINN/ (Spanish Ministry of 
Science and Innovation) - FEDER through grants AYA2009-06934, AYA2009-14648-C02-01 
and CONSOLIDER CSD2007-00050.  \end{acknowledgements}

\bibliographystyle{aa} % style aa.bst

\end{document}